\documentclass[a4paper]{jpconf}
\usepackage{graphicx}

\newcommand{\bea}{\begin{eqnarray}}
\newcommand{\eea}{\end{eqnarray}}

\newcommand{\be}{\begin{equation}}
\newcommand{\ee}{\end{equation}}

\newcommand{\K}{\mathbf{k}}

\newcommand{\R}{\mathbf{r}}

\newcommand{\si}{\mathbf{S_i}}

\newcommand{\sj}{\mathbf{S_j}}

\begin{document}

\title{Magnetization plateaux in the classical Shastry-Sutherland lattice}

\author{M Moliner$^1$, D C Cabra$^{1,2,3}$, A Honecker$^4$, P Pujol$^5$ and F Stauffer$^6$}

\address{$^1$ Laboratoire de Physique Th\'eorique, Universit\'e Louis Pasteur, 3 rue de l'Universit\'e, F-67084 Strasbourg Cedex, France}
\address{$^2$ Departamento de F\'{\i}sica, Universidad Nacional de la Plata, C.C.\ 67, (1900) La Plata, Argentina}
\address{$^3$ Facultad de Ingenier\'\i a, Universidad Nacional de Lomas de Zamora, Cno. de Cintura y Juan XXIII, (1832) Lomas de Zamora, Argentina.}
\address{$^4$ Institut f\"ur Theoretische Physik, Universit\"at G\"ottingen,
 37077 G\"ottingen, Germany}
\address{$^5$ Laboratoire de Physique Th\'eorique, IRSAMC, UPS and CNRS, Universit\'e de Toulouse, 118, route de Narbonne, F-31062 Toulouse, France}
\address{$^6$ RBS Service Recherche \& D\'eveloppement 11 rue Icare Strasbourg - Entzheim 67836 Tanneries Cedex, France}

\ead{moliner@lpt1.u-strasbg.fr}
\begin{abstract}
We investigated the classical Shastry-Sutherland lattice under an external magnetic field in order to understand the recently discovered magnetization plateaux in the rare-earth tetraborides compounds RB$_4$. 
A detailed study of the role of thermal fluctuations was carried out by mean of classical spin waves theory and Monte-Carlo simulations. Magnetization quasi-plateaux were observed at $1/3$ of the saturation magnetization at non zero temperature. We showed that the existence of these quasi-plateaux is due to an entropic selection of a particular collinear state. We also obtained a phase diagram that shows the domains of existence of different spin configurations in the magnetic field versus temperature plane.
\end{abstract}
\begin{figure}[h]
 \begin{minipage}{12pc}
  \begin{center}
   \includegraphics[width=10pc]{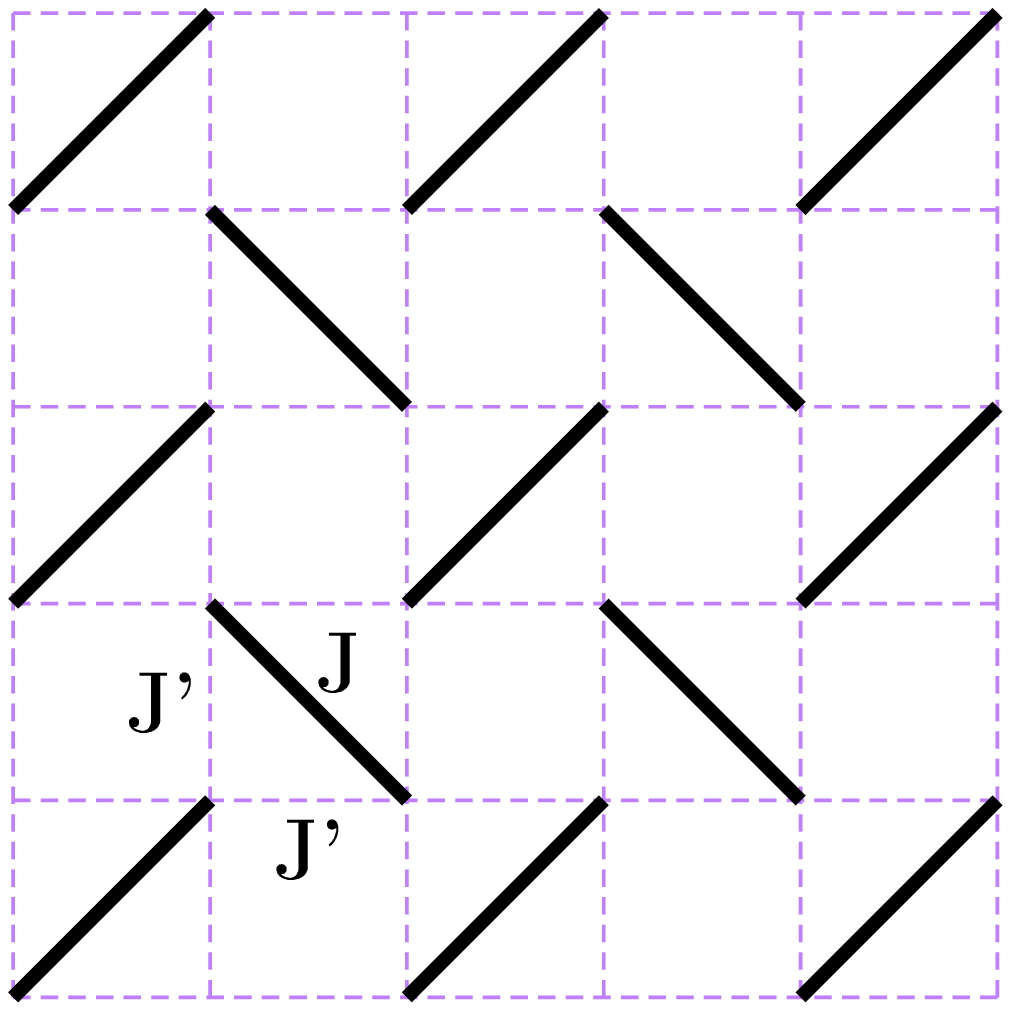}
  \end{center}
  \caption{\label{fig:SSL}The SSL with magnetic couplings $J'$ bonds along the edges of the squares and $J$ along the diagonals.}
 \end{minipage}\hspace{2pc}
 \begin{minipage}{10pc}
  \begin{center}
   \includegraphics[width=9pc]{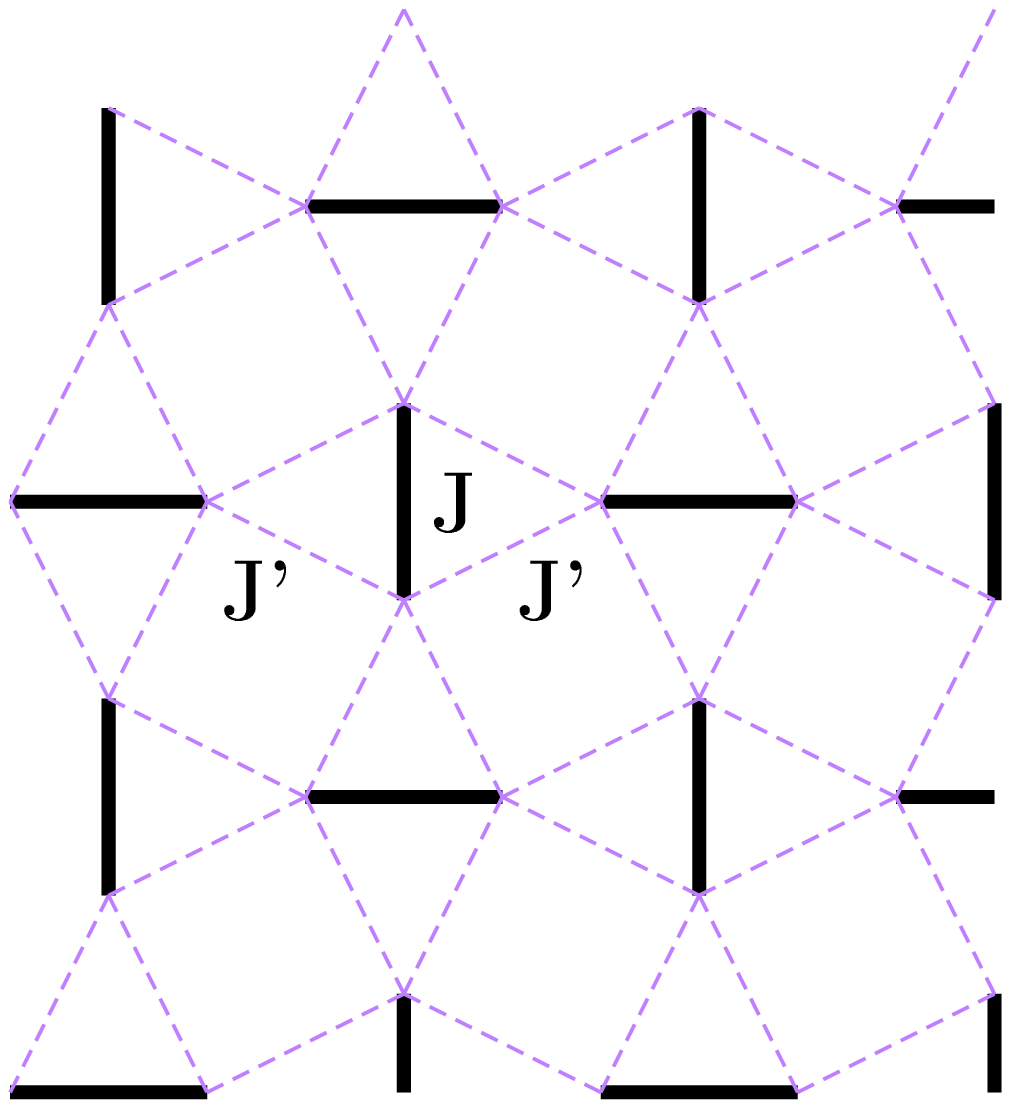}
  \end{center}
  \caption{\label{fig:SSL_real}The topologically identical structure realized in SrCu$_2$(BO$_3$)$_2$ and in RB$_4$.}
 \end{minipage}\hspace{2pc}
 \begin{minipage}{11pc}
  \begin{center}
  \includegraphics[width=10pc]{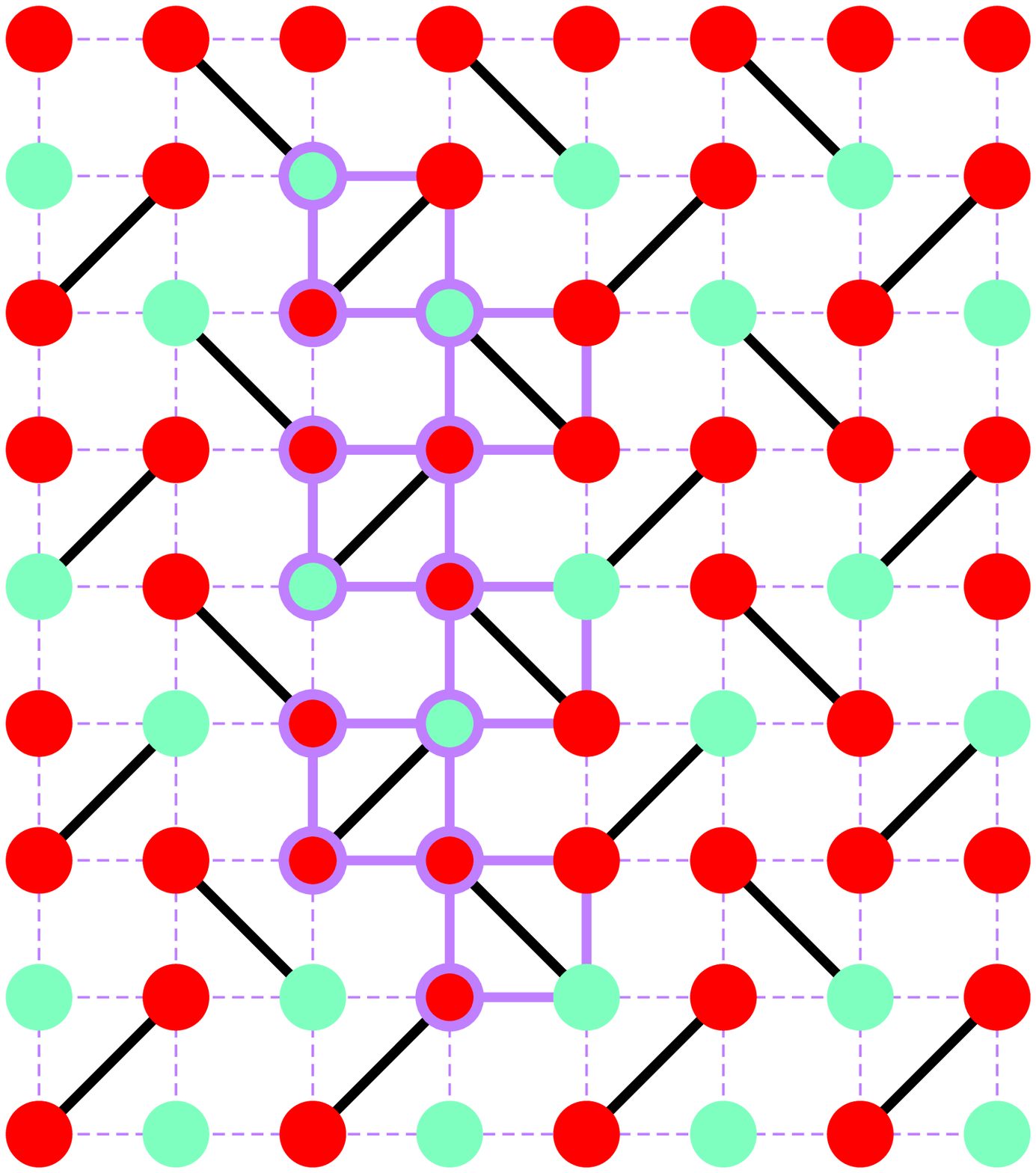}
  \end{center}
  \caption{\label{fig:UUD}The $UUD$ state at $M/M_{sat}=1/3$ and the 12 spins unit cell (bold lines). Each triangle contains two spins up (red) and one spin down (blue).}
 \end{minipage} 
\end{figure}

The Shastry-Sutherland lattice (SSL) was first considered in the early eighties by Shastry and Sutherland \cite{SS}. This frustrated lattice can be described as a square lattice with antiferromagnetic couplings $J'$ and additional diagonal antiferromagnetic couplings $J$ in one square over two (figure \ref{fig:SSL}). The quantum version of the SSL was widely studied both theoretically and experimentally after its experimental realization in the SrCu$_2$(BO$_3$)$_2$ compound. The quantum SSL was found to present a rich phase diagram as the ration $J'/J$ varies. The existence of magnetization plateaux was theoretically predicted and experimentally verified for various rational values of the saturated magnetization $M/M_{sat} = 1/8$, $1/4$, $1/3$ and $1/2$ \cite{reviewSrCu2BO3}.\\ 
Recently magnetization plateaux were discovered in rare-earth tetraborides RB$_4$ in which the rare-earth ions are placed in the $(001)$ plane according to a lattice that is topologically equivalent to the SSL (figure \ref{fig:SSL_real}). RB$_4$ compounds present large total angular momenta that justify a \emph{classical} description of the SSL. 
High magnetic-fields experiments revealed plateaux in the magnetization curves of TbB$_4$ at $M/M_{sat}=1/4, 1/3$ and $1/2$ \cite{TbB4}, in ErB$_4$  at $M/M_{sat}=1/2$  \cite{ErB4} and in TmB$_4$ $M/M_{sat}=1/2$ and $M/M_{sat}=1/8$ \cite{TmB4}.\\
We study Heisenberg spins on the SSL in the classical limit. The spins are represented by vectors $\si$ with $|\!|\si|\!|=1$. The magnetic field is applied
to the $z-$axis.
The Hamiltonian is given by:
\be
 \mathcal{H} =  \frac{J'}{J} \sum^N_{edges}\!\! \si.\sj
              + \!\!\sum^N_{diagonal}\!\!\!\!\! \si.\sj
              - \frac{h}{J} \sum^N_{i} S_i^z
\label{Hamiltonian}
\ee
The sums are taken over the edges of the squares with antiferromagnetic couplings $J'$ and over the diagonals with $J$ couplings. 
In the absence of applied magnetic field the lowest energy configuration (LEC) is coplanar. It is N\'eel ordered for $J'/J \ge 1$ else it is a spiral state with an angle $\varphi = \pi \pm \arccos\big(\frac{J'}{J}\big)$ between nearest-neighbour spins. Those two possible values for $\varphi$ create a supplementary discrete degeneracy above the continuous one \cite{SS}.
For the particular ratio of magnetic couplings $J'/J=1/2$, the Hamiltonian Eq. (\ref{Hamiltonian}) can be written as a sum on triangles up to a constant term: 
\bea
  \mathcal{H}_{\Delta} = \frac{1}{J}\sum_{\Delta}^{N_{\Delta}}\Big( \frac{J'}{2} \mathbf{S_{\Delta}}^2 - \frac{h}{3} \mathbf{S_{\Delta}} \Big)
 \mbox{ with $J'/J = 1/2$ }
 \label{Hamiltonian_triangles}
\eea
Where $\mathbf{S_{\Delta}}=\sum_{i\in \Delta} \si$ is the total magnetization of one triangle  and $N_{\Delta}$ is the number of triangles ($N_{\Delta}=N$). 
Minimizing the energy on a single triangle, one obtains the constraint:
\be
  S_{\Delta} = \frac{h}{3J'} \mbox{ with $J'/J = 1/2$}
 \label{plaquette_constrain}
\ee
The LEC is obtained when this constraint is satisfied in every triangle. In the absence of magnetic field it is a coplanar $120\,^{\circ}$ structure with a 12 spins unit cell.
Saturation is reached at $h_{sat}=9J'$ while the magnetic field at $M/M_{sat}=1/3$ is $h_{1/3} = 3J'$ (with $J'/J = 1/2$). At $h=h_{1/3}$ different spin configurations have the same classical energy: for example, the "umbrella configuration" in which the three sublattices always have the same projection on the $z-$axis for all magnetic field applied and the "Up-Up-Down" ($UUD$) state. In this latter configuration each triangle contains two spins up and one spin down (figure \ref{fig:UUD}). The existence of a classical plateaux requires the LEC to be $UUD$ \cite{HidaAffleck}. Since energetic considerations do not favour the $UUD$ state over the other configurations no magnetization plateaux are expected at zero temperature.\\
At finite temperature the most favourable configuration is also determined by the density of states above the LEC. Under thermal fluctuations the classical spin wave spectrum of a particular state (or subset of states) can present \emph{soft modes} whose energy in spin deviations is quartic in state of quadratic. Those modes contribute less to the free energy and the state with more soft modes is favoured. This phenomenon is known as \emph{Order by Disorder} \cite{ObD}.\\
We computed the spin-wave spectrum of the $UUD$ state. The coordinates of the 12 spins of the unit cell are rewritten
\be
 \si (\R_i) = \Big(\epsilon^x_i (\R_i),\epsilon^y_i (\R_i),1- 1/2\big((\epsilon^x_i)^2 + (\epsilon^y_i)^2 \big) \Big)
 \label{spin_T}
\ee
Up to second order in spin deviations applied on the 12 sublattices the Hamiltonian reads: 
\be
 \mathcal{H} =   E_0 
               + \sum_{\K,v=x,y}\!\!\! \mathcal{V}_v^t(-\K) \mathcal{M} \mathcal{V}_v(\K) 
               +  \sum_{n > 2} \mathcal{H}_n 
 \label{Hfluct}         
\ee
Where $E_0$ is the classical energy and $\mathcal{H}_n \sim \mathcal{O}(\epsilon^n)$ and 
$\mathcal{V}_v^t(-\K) = \Big(\tilde{\epsilon}^v_{1}(-\K), \dots ,\tilde{\epsilon}^v_{12}(-\K) \Big)$. 
Eigenvalues of the matrix $\mathcal{M}$ should be determined by solving an order 12 polynomial which could not be done analytically. Its determinant is given by: 
\be
 Det(\mathcal{M}) \propto \big( -2 + 3\cos k_x - 3\cos 2k_x  + \cos 3k_x  + \cos k_y  \big)^2 
 \label{determinant}
\ee 
The expression for the determinant shows that \emph{lines} of soft modes can be found. However since the dimension of this submanifold is only 1, in the thermodynamic limit, those lines of soft modes no longer play a role and they will not create a fall in the specific heat \cite{Modescounting}.
In a similar way as in the classical triangular lattice, in the SSL, the $UUD$ state is selected by entropy selection without any full branch of soft modes, in contrast to the kagome case \cite{KagomeZhitomirsky}.\\
We performed MC simulations using the standard Metropolis algorithm on square samples of size $L=\sqrt{N}=6,12,18,24$ and $30$ with periodic boundary conditions. System sizes are chosen so as to be commensurable with the 12 spins unit cell represented in figure \ref{fig:UUD}. 
Magnetization, susceptibility $\chi = dM/dh$ and specific heat $C_h$ were computed. The thermal equilibrium was reached by collecting data with up to $\sim 10^{7}$ MC steps on an average number of five simulations per point.\\
Quasi-plateaux are observed in the magnetization curves at $M/M_{sat}=1/3$. Their width increases with the system size which confirms the fact that the its stabilization is not due {\it only} to the lines of soft modes. 

\begin{figure}[h]
 \begin{minipage}{18pc}
  \begin{center}
   \includegraphics[width=18pc]{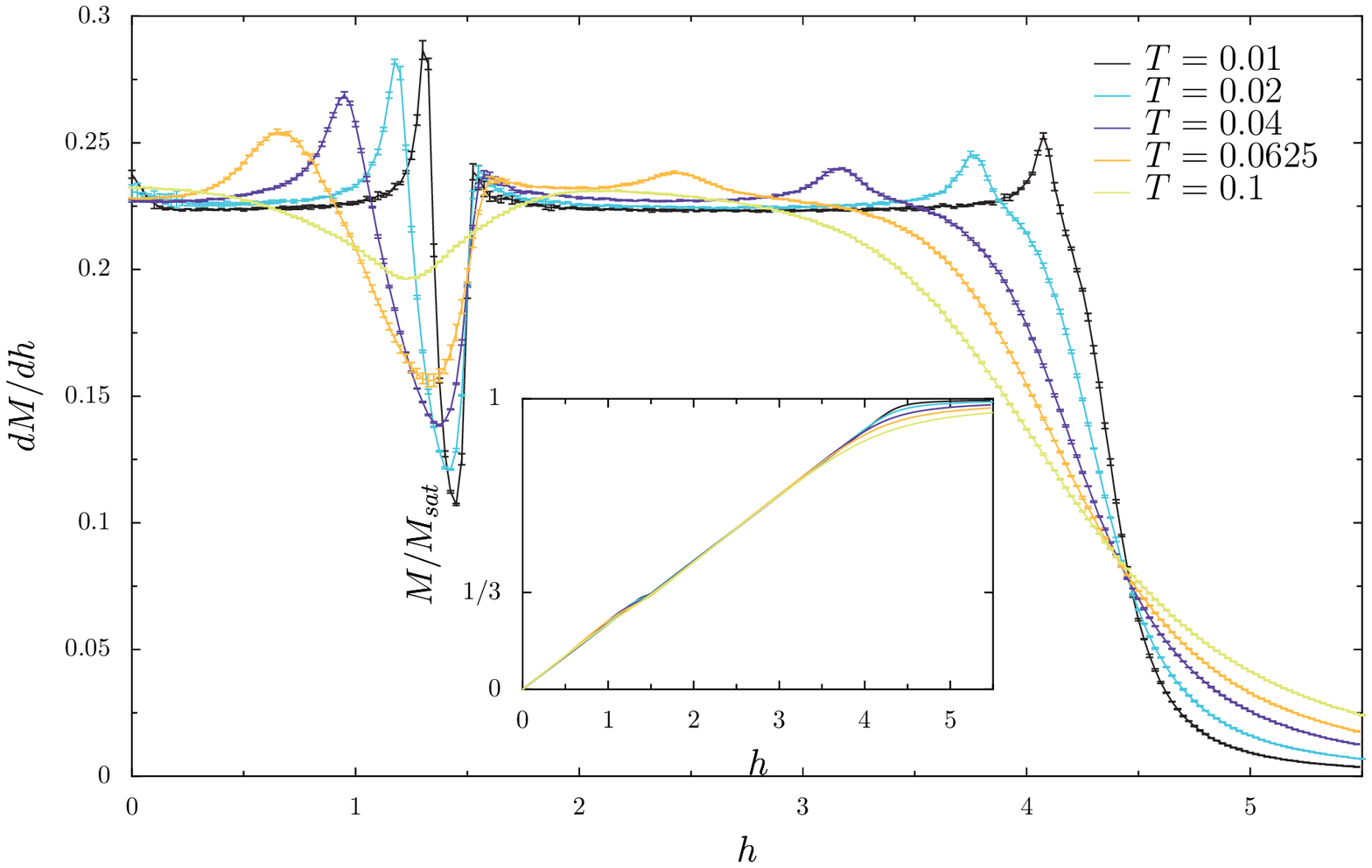}
  \end{center}
  \caption{\label{fig:R05Teffect} Susceptibility and magnetization (inset) for a $L=18$ system at $J'/J=1/2$.}
 \end{minipage}\hspace{2pc}
 \begin{minipage}{18pc}
  \begin{center}
   \includegraphics[width=18pc]{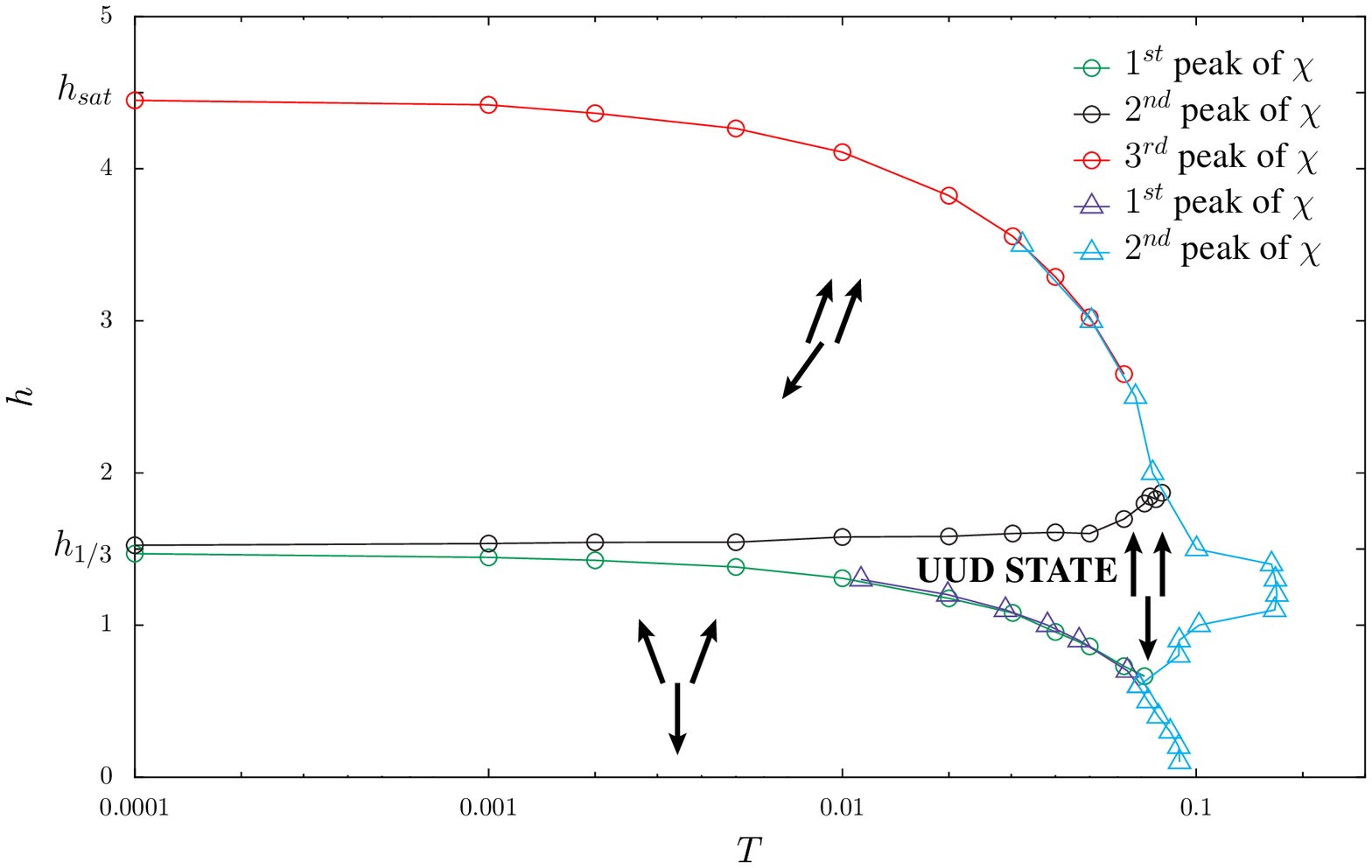}
  \end{center}
  \caption{\label{fig:R05_phase_diag}Phase diagram of the classical SSL at $J'/J=1/2$.}
 \end{minipage}\hspace{2pc}
\end{figure}
Figure \ref{fig:R05Teffect} shows the effect of the temperature on the quasi-plateaux. As the temperature decreases the quasi-plateaux tend to disappear which is in agreement with the analytical prediction. Their width increases with temperature until they are destroyed by temperatures higher than $T \sim 0.1$. The susceptibility presents two very sharp peaks around $h_{1/3}$ that become rounded and disappear as the temperature increases. 
The positions of those peaks were studied in order to determine qualitatively the domain on existence of the collinear phase in the $(h,T)$ plane (see below). We performed MC simulations on $L=12$ systems in order to obtain the schematic phase diagram presented in figure \ref{fig:R05_phase_diag}. Numbers are attributed to the susceptibility peaks from low to high temperature in magnetic field scan (curves with circles) and from low to high magnetic field in temperature scan (curves with triangles).\\ 
This phase diagram presents similarities with the ones of the classical the Kagom\'e \cite{KagomeZhitomirsky} and triangular lattices \cite{CTL}. 
In the high temperature region ($ \sim T >0.1$) the system enters the disordered paramagnetic phase. The low temperature region contains three phases. The nature of the phase transitions is still to be determined \cite{us}. The collinear $UUD$ phase is the LEC for values of the magnetic field below and until $h_{1/3}$. As the temperature tends to zero this phase width converges to a single point at exactly  $h_{1/3}=3/2$. The phases below and above $UUD$ contain coplanar states for which quasi-long range order is expected. In the low field region each triangles contains one spins below the $(x,y)$ plane while the remaining two spins raise until the $UUD$ state. In high field region the down spin raises until saturation ($UUU$ state).\\
We also studied the effect of a small modification $\epsilon$ of the magnetic coupling ratio $J'/J=1/2$. Magnetization plateaux are also observed for $J'/J=1/2 \pm \epsilon$. $UUD$ is no longer the LEC at "very low temperature" (i.e. accessible with lowest-order thermal fluctuations). In state the LEC is a non-collinear umbrella. However at higher temperature $UUD$ becomes the LEC and there still exists a quasi-plateau phase in the $(h,T)$ plane surrounded by the same not collinear phases as in the case  $J'/J=1/2$.\\ 
In conclusion, we found magnetization quasi-plateaux at $M/M_{sat}=1/3$ in the classical SSL and showed that they are due to the entropic selection of the collinear $UUD$ state. A phase diagram, that reminds the one of the classical triangular lattice,  was established in the $(h,T)$ plane for the particular coupling ratio $J'/J=1/2$. We also showed that, because of an entropic selection of the collinear "UUD"state, those quasi-plateaux survive under small variations $\epsilon$ of the magnetic coupling around $J'/J=1/2$ which disfavor energetically this collinear configuration.


This work was partially supported by ESF (grants 1872 and 2268) and the DFG (grant HO~2325/4-1).
DCC, AH and PP would like to thank M~E~Zhitomirsky for collaboration
during an early stage of this work.


\end{document}